# The Preliminary Results on Analysis of TAIGA-IACT Images Using Convolutional Neural Networks

**Elizaveta Gres**[a,*] **and Alexander Kryukov**[b]

[a]*Research Institute of Applied Physics, Irkutsk State University,*
 *20 Gagarina Blvd, Irkutsk, Russian Federation*

[b]*Skobeltsyn Institute of Nuclear Physics, M.V. Lomonosov Moscow State University,*
 *1(2) Leninskie gory, Moscow, Russian Federation*

 *E-mail:* greseo@mail.ru, kryukov@theory.sinp.msu.ru

The imaging Cherenkov telescopes TAIGA-IACT, located in the Tunka valley of the republic Buryatia, accumulate a lot of data in a short period of time which must be efficiently and quickly analyzed. One of the methods of such analysis is the machine learning, which has proven its effectiveness in many technological and scientific fields in recent years. The aim of the work is to study the possibility of the machine learning application to solve the tasks set for TAIGA-IACT: the identification of the primary particle of cosmic rays and reconstruction their physical parameters. In the work the method of Convolutional Neural Networks (CNN) was applied to process and analyze Monte-Carlo events simulated with CORSIKA. Also various CNN architectures for the processing were considered. It has been demonstrated that this method gives good results in the determining the type of primary particles of Extensive Air Shower (EAS) and the reconstruction of gamma-rays energy. The results are significantly improved in the case of stereoscopic observations.

[*]Speaker

## 1. Introduction

To create a more complete picture of the evolution of the Universe and check the adequacy of theoretical models, it is necessary to investigate the interactions of elementary particles in the energy range above 100 TeV to study the phenomena in the Universe [1]. The main objective of the gamma-ray astronomy is the identification and research of high-energy gamma radiation sources. Such objects include supernova remnants, active galactic nuclei, and much more. Measuring the flux, energy spectrum and direction of arrival of gamma photons helps to find answers regarding the generation mechanism of high energy gamma radiation and the morphology of sources.

At the moment, gamma radiation can be observed both from space and from the surface. The registration of gamma quanta with energies of TeVs is several orders of magnitude lower than photons with lower energies; therefore, they are registered using ground-based installations. Ground-based Imaging Atmospheric Cherenkov Telescopes (IACTs) are the main instruments for observation of the high-energy gamma radiation. These telescopes register not the gamma quanta (or cosmic rays) themselves, but the Cherenkov radiation arising in the process of the Extensive Air Shower (EAS) generated by them.

TAIGA-IACT is a part of the hybrid installation TAIGA (Tunka Advanced Instrument for cosmic ray physics and Gamma-ray Astronomy), located in the Tunka valley of the republic Buryatia [2]. These telescopes have large spherical segmented mirrors with a camera in the focus of the mirrors. The cameras contain a matrix of 560-590 photomultipliers (PMT). The main task of the TAIGA-IACT is to separate gamma events from the cosmic ray background and reconstruct the parameters of the primary particle.

One of the standard methods for image processing obtained by IACTs is the Hillas parameter method [3]. The essence of this method is that the spot in the camera is described by an ellipse with certain parameters, according to which the classification and restoration of events are carried out. At the moment the use of convolutional neural networks [4] as one of the methods of machine learning for TAIGA-IACT image processing has not been implemented to real data, therefore this work allowed us to study the prospects of using this method. It is known that other IACT installations [5, 6] have shown promising results in image analysis of model data using convolutional neural networks. CNNs were also used for the TAIGA-IACT model data of one telescope [7, 8]. However an imbalance of particle fluxes is observed in an experiment, thus it is necessary to consider the classification of events in the case of an unequal ratio of gamma quanta and hadrons. TAIGA-IACT also consists of two telescopes [2], so the CNN method will provide the estimation the quality of event energy reconstruction in the case of joint observations by several telescopes.

The method of convolutional neural networks which was applied for processing of Monte-Carlo events is presented in this article. Several convolutional neural network structures have been developed, trained and tested. The quality of event classification and event energy reconstruction were evaluated. The case of observations of joint gamma events with two telescopes (stereo-mode) was also considered and compared with the observations with one telescope (mono-mode). Results have been demonstrated that CNNs improve the selection (around 100 times) for unbalanced ratio of gamma quanta and hadrons compared to the equal ratio. Also increasing the number of telescopes during observations linearly improves the accuracy of determining the energy of events.

## 2. Method of convolutional neural networks

Neural networks are mathematical representations of neurons work in the brain [4]. So, in one neuron the weighted signals coming from the previous neurons are summed up. After that the neuron generates an output signal through the activation function. Neurons multilayered form a neural network. During the training with some examples weights are adjusted between neurons through the backpropagation method [9] which is an analogue of the gradient descent method. This method allows to reduce the error between the predicted by network and the true result through the calculation of derivatives.

Convolutional neural networks (CNN) appeared as a result of studying the visual cortex of the brain [4]. It was found out that in the visual cortex there is a small local receptor field that reacts to visual irritants located in a limited area of the visual field. This led to the emergence of a new structure in neural networks – convolutional layers. Convolutional layers use a small-sized weight matrix ("local receptor field"), which is called a filter or kernel. With its help a sequential "scan" of the image takes place through the convolution operation. This operation allows to identify common structures and features in the images regardless of their location, on the basis of which ordinary neurons allocate the necessary response. Due to this CNN is one of the best ways to analyze images. CNNs also help to solve several data processing tasks (in our case, classification and regression) with minimal changes in the structure, for example, by changing only the activation function.

### 2.1 Used CNN architectures

In this work the programming of convolutional neural networks was carried out in Python using a special Tensorflow library together with Keras [10]. Schematic images of the network architectures developed during debugging are shown in Figure 1. During the debugging the numerical values of some hyperparameters of neural networks (such as the learning rate, dropout chance) were determined. As can be seen from the figure, two or three convolutional layers depending on the task were used in the models. The regression problem had to use more convolutional layers compared to the classification network. This is due to the fact that the regression problem requires a more thorough analysis of the image, since the image depends on many factors: the type of primary particle, the distance to the telescope, and other factors. Also, the main differences between these networks are related to the output value due to changes in the activation function and the method of calculating the error. Mean squared error calculation was used for the regression task, while binary cross entropy loss calculation were applied in classification network. Dropout and Pooling layers helped with the overfitting problem [4].

Along with user developed user structures the architectures of well-known networks were also studied: ResNet и GoogLeNet. These networks showed one of the best results in ImageNet Large Scale Visual Recognition Challenge [11]. For comparison they were reduced in such a way that the number of trained parameters approximately coincided with the number of parameters in user networks. Their simplified structure is also shown in Figure 1.

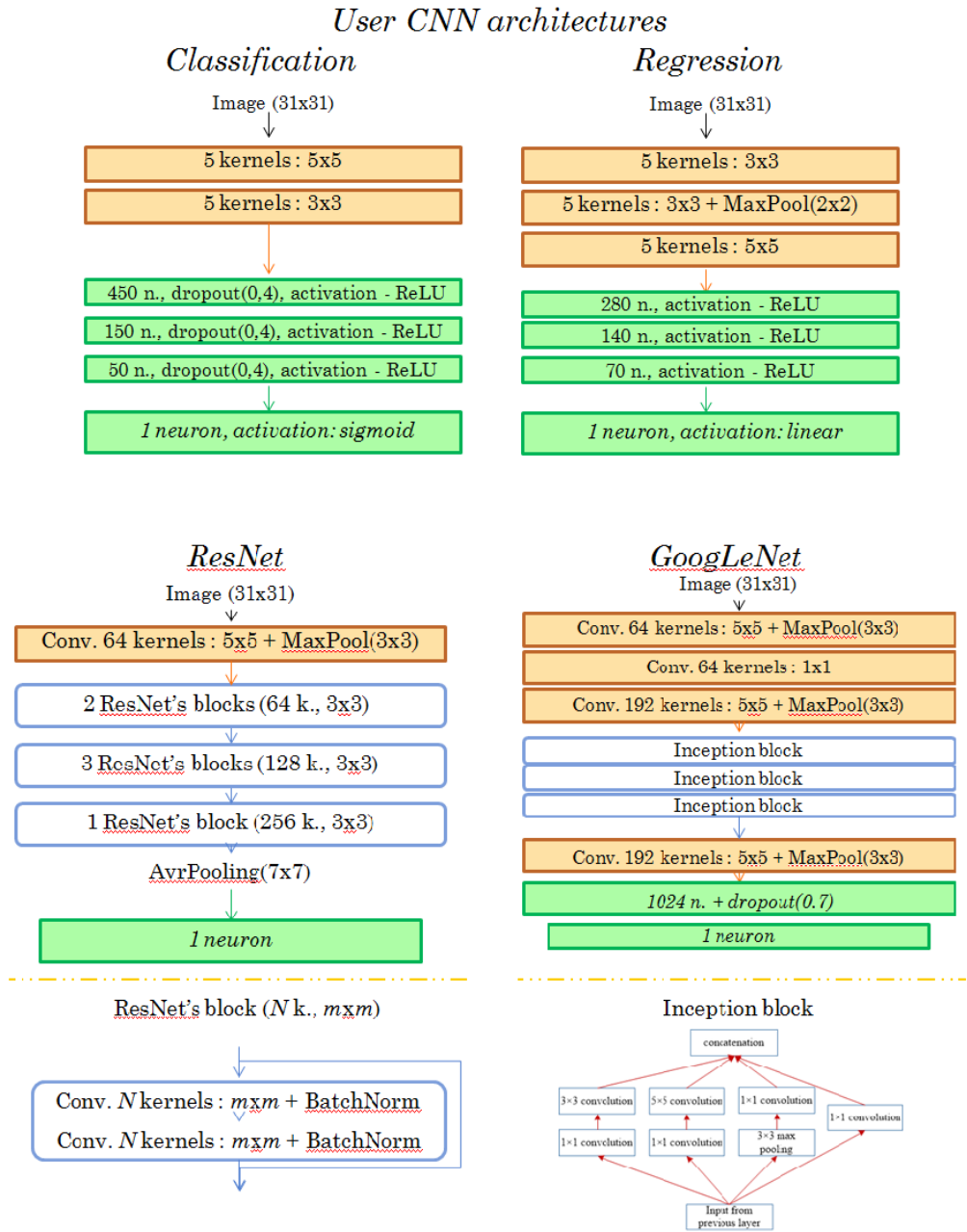

**Figure 1:** The CNN architectures used in work on processing and analysis of simulated Monte-Carlo events of TAIGA-IACT

## 3. Model data

The training and testing of neural networks was carried out on three different sets of Monte-Carlo events simulated with CORSIKA [12]. A description of each set is presented in Table 1. Set №1 and №2 were used for classification and regression, while the third set was only applied to study the quality of energy reconstruction in stereo-mode. It should be noted that due to the small number of events in the first set, its artificial expansion were occurred by rotating the images relative to the center of the camera by every 60 degrees. Thus, the first set was expanded in 6 times.

| Set | Total events (gamma/proton) | Train and validation ratio | Energies |
|---|---|---|---|
| 1 | 30 000 (17 500 / 12 500) | 2:1 | Hadron: 2-100 TeV γ: 1-60 TeV |
| 2 | 200 000 (100 000 / 100 000) | 3:1 | Hadron: 5-100 TeV γ: 2-50 TeV |
| 3 | 18 000 (only gamma) | 2:1 | 1-50 TeV |

**Table 1:** The description of Monte-Carlo data

The following steps were performed as a preprocessing of images to improve training. Single pixels were removed during the image cleaning process. Then each image was squared by shifting the pixels relative to each other [13]. Images with a size of 31x31 pixels were obtained as a result of this transformation. Also the amplitude of each pixel $x_i$ in each image was scaled as follows:

$$\tilde{x}_i = \frac{1}{9} \ln(1 + x_i), \qquad (1)$$

where $i$ is the pixel number in the image, $\tilde{x}_i$ is the scaled pixel amplitude. The number 9 limits the change of $\tilde{x}_i$ in the range of values from 0 to 1, since the maximum value of the pixel amplitude in the training and validation set is unknown in advance.

## 4. The study of the obtained results

As mentioned above, the main objectives of the TAIGA-IACT are:
- The selection of gamma-ray events from the hadron background (classification);
- The restoration of parameters of the primary particle (regression), in particular, the energy.

The results of using CNN to solve these problems are presented further in the article.

### 4.1. The classification task

Training and validation were applied on the first and second sets. The accuracy of determining the event class was 95-96% regardless of the set. This result showed the independence of the rotated images during CNN operation. Due to the peculiarities of the filters of convolutional networks this result was expected [4, 13].

To estimate the classification and subsequent comparison, the quality parameter of selection $Q$ was considered, which can be defined as the ratio of the significance criterion $S$ before and after selection by the neural network (the significance criterion in modeling determines how many times the expected signal exceeds the background):

$$Q = \frac{S_{after}}{S_{before}} = \frac{N_g}{\sqrt{N_g + N_{hg}}} \bigg/ \frac{N_{gammas}}{\sqrt{N_{total}}}, \qquad (2)$$

where $N_g$ – the number of true gamma events identified by the CNN as gammas, $N_{hg}$ – the number of proton events identified by the CNN as gammas events, $N_{gammas}$ – the total number of gamma events in the set, $N_{total}$ – the number of all events in the set. Since the CNN classification gives the probability that the event is a gamma quantum the class separation threshold was defined in such a way that approximately 50% of the true gamma quanta were determined correctly. Therefore, the threshold was approximately 0.97.

It is known [14] that the fluxes of gamma quanta and hadrons differ greatly (approximately 1:10 000) in a real experiment, so the parameter $Q$ was calculated in the case of different ratios of gamma photons and hadrons. The results of calculating the quality parameter of selection are demonstrated in Table 2. The table shows that there is no improvement in the quality of classification with an equal ratio of gamma quanta and protons. But there is a good suppression of proton events with unequal ratio, but the significance $S$ becomes small (1 sigma). Among the various structures of convolutional networks GoogLeNet gives the best result.

| The ratio of gamma and hadrons | Applied CNN architectures | $S_{after}$ | $Q$ |
|---|---|---|---|
| 1:1 | User CNN | 107.24 | 1.07 |
| 1:100 | User CNN | 3.69 | 4.11 |
| 1:1000 | User CNN | 0.92 | 5.04 |
| 1:1000 | ResNet | 1.04 | 5.72 |
| 1:1000 | GoogLeNet | 1.13 | 6.21 |

**Table 2:** Data classification with different class balance

**4.2 The regression task**

In the case of regression the evaluation measure was the relative error in determining the energy $\delta$ which is defined as follows:

$$\delta = \frac{|E_{pred} - E_{true}|}{E_{true}}, \qquad (3)$$

where $E_{pred}$ – the energy predicted by CNN, $E_{true}$ – the true energy value. In this task the CNNs tried to restore the energy of events in the case of a mixed set (there are gamma quanta and hadrons in the set), and in the case of a set of only gamma quanta. Figure 2 shows the relative error distributions (denoted as *rel_err* on the graph) for both cases. Thus, it is demonstrated that the median value of the relative error for the mixed set is 32%, while the "clear" set is 23-25%. At the same time it can be seen that different CNN structures does not greatly improve the result, by no more than 3%. The best result in determining energy is given by the GoogLeNet.

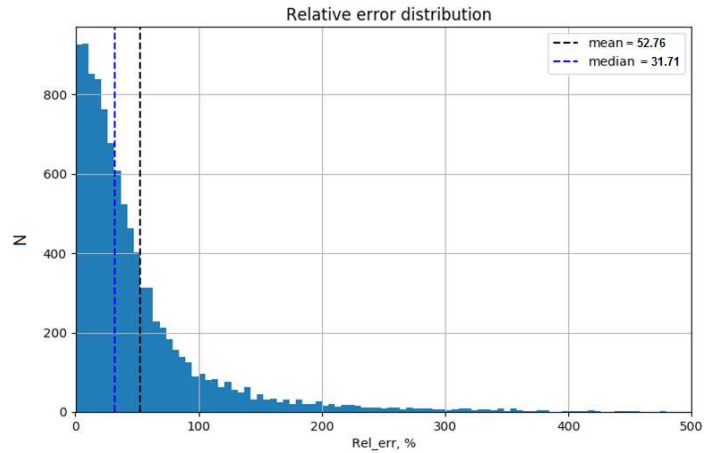

a)

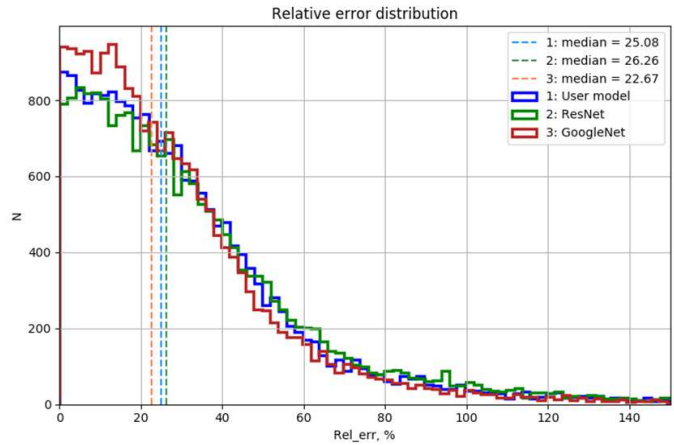

b)

**Figure 2**: The distribution of relative error of energy reconstruction in case of mixed (gammas and protons) events (a) and in case of only gamma-photons (b)

The study of event energy recovery in the case of the stereo mode and comparison with the mono mode was carried out only for gamma events and using the user CNN model. The second telescope was taken into account when modifying the network as follows. In the regression user architecture an additional input has been added with the same number of convolutional layers as the first input (see Fig. 1). After that the two inputs, or channels, were combined using dense layers.

The results of the evaluation of the energy determination and comparison with the mono-mode are presented in Figure 3. As can be seen in the figure, stereoscopic observations double the accuracy of energy recovery. Thus, the relative error decreased from 23% to 14%.

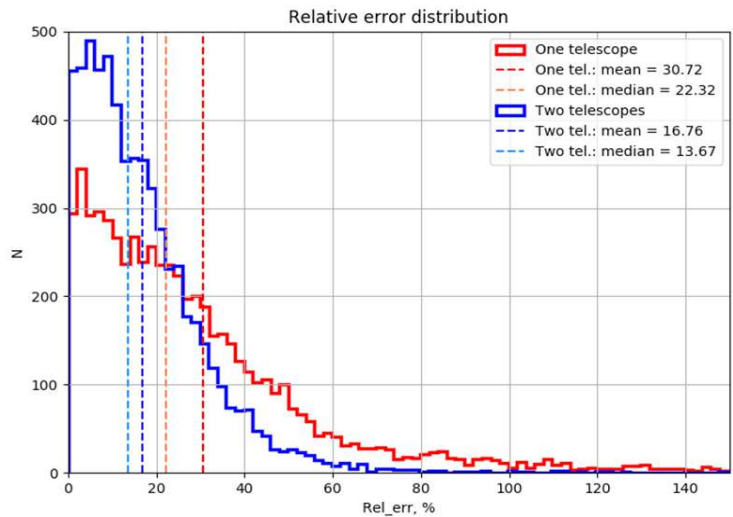

**Figure 3**: The distribution of relative error of energy reconstruction for stereo- and mono-modes of TAIGA-IACT for gamma-photons events

## 5. Conclusion

For ground-based TAIGA-IACTs there is a problem of reliably determining the type of recorded events, as well as the problem of restoring the initial parameters of particles that generate EASs. The method of convolutional neural networks was applied to solve these problems.

The results demonstrated that CNN classification suppresses the proton background greatly (around 100 times), but significance is low (around 1 sigma). Energy reconstruction showed the around 24% relative error for one telescope, and 14% – for two telescopes. ResNet and GoogLeNet demonstrated a slight results improvement in both particle type and energy determination.

Thus, in perspective this method can be used for the energy restoration, as it gives good results. Also CNN for good background suppression can be considered as additional selection threshold

## Acknowledgements

The authors would like to thank the TAIGA collaboration for their support and provision of the model data.